\documentclass[prl, aps, showpacs,tightenlines, twocolumn, amsmath,amssymb]{revtex4}
\usepackage[dvips]{graphicx,color}
\usepackage{dcolumn}
\usepackage{bm}

\newcommand\beq{\begin{equation}}
\newcommand\eeq{\end{equation}}
\newcommand\beqa{\begin{eqnarray}}
\newcommand\eeqa{\end{eqnarray}}

\newcommand{\ds}[1]{#1 \hspace{-0.5em}/}  

\newcommand\bgamma{\mbox{\boldmath$\gamma$}}

\newcommand\bq{{\bf q}}

\newcommand\bx{{\bf x}}

\newcommand\bk{{\bf k}}

\newcommand\balpha{\mbox{\boldmath$\alpha$}}
\newcommand\bP{{\bf P}}
\newcommand\by{{\bf y}}

\begin{document}

\title{Novel Lifshitz point for chiral transition in the magnetic field}%

\author{Toshitaka Tatsumi} 
\email{tatsumi@ruby.scphys.kyoto-u.ac.jp}
\author{Kazuya Nishiyama}
\author{Shintaro Karasawa}
\affiliation{%
Department of Physics, Kyoto University, Kyoto 606-8502, Japan
}%

\date{\today}

\begin{abstract}
Based on the generalized Ginzburg-Landau theory, chiral phase transition is discussed 
in the presence of magnetic field. Considering the chiral density wave we show 
chiral anomaly gives rise to an inhomogeneous chiral phase for nonzero 
quark-number chemical potential. Novel Lifshitz point appears on the vanishing chemical potential line, 
which may be directly explored by the lattice QCD simulation.
 \end{abstract}

\pacs{ 11.30.Rd,  
 12.38.Gc,
 25.75.Nq, 
}
\maketitle


One of the recent development for the QCD phase diagram may be a possible formation of inhomogeneous chiral phases and their implications on high-energy heavy-ion collisions or compact stars \cite{revqcd}. They are specified by the spatially inhomogeneous chiral condensates and quite similar to FFLO state in superconductivity \cite{fflo} or the textured phase in magnetism \cite{tex}. Similar subject has been also discussed in the context of color superconductivity 
\cite{alf}. Considering spatial modulation of the $\bar q q$ condensates in quark matter, 
they take form, 
$
\langle{\bar\psi}\psi\rangle+i\langle{\bar\psi}i\gamma_5\tau_3\psi\rangle
\equiv \Delta(\bx){\rm exp}
\left( i\theta(\bx)\right),
$ 
within $SU(2)_L\times SU(2)_R$ chiral symmetry. Various types of the condensates can be considered: two kinds of one-dimensional order are well known in 1+ 3 dimensions within the Nambu-Jona-Lasinio (NJL) model: one is called dual chiral density wave (DCDW) characterized by the uniform amplitude $\Delta$ and $\theta=\bq\cdot \bx$ \cite{nak}, and the other is called real kink crystal (RKC) by the spatially periodic function of $\Delta$ without $\theta$ \cite{nic}. 
These configurations can be also obtained by embedding the Hartree-Fock solutions in the 1+1 dimensional models;
the general form of the condensates has been found through the studies of the phase structure of the NJL$_2$ model or Gross-Neveu (GN) model 
\cite{dun}. Similar subject has been also discussed in quarkyonic matter \cite{koj}.

In this Letter we consider the DCDW-type configuration specified by $\Delta(\bx)$ and $\theta(\bx)$.  
Non-vanishing $\Delta$ implies spontaneous symmetry breaking (SSB) of chiral symmetry. Then we can easily observed that the state can be described by operating 
the {\it local} chiral rotation with the chiral angle $\theta(\bx)$, $U_{\rm DCDW}(\theta(\bx))={\rm exp}\left[i\int\theta(\bx)A_3^0(\bx)d^3x\right]$, on the quark matter with $\Delta$, $|{\rm DCDW}\rangle=U_{{\rm DCDW}}|{\rm QM};\Delta\rangle$, where $A_3^\mu(\bx)$ is the isospin 3-rd component of the axial-vector current; consider the Dirac Hamiltonian,
$H_D^0=-i\balpha\cdot\nabla+\gamma^0m(\bx)$, where we assume that the mass function $m(\bx)$ is given by the scalar condensate, $m(\bx)=-2G\langle{\bar\psi}\psi\rangle$ as in the NJL-like models, ${\cal L}={\bar\psi}i\ds{\partial}\psi
+G({\bar\psi}\psi)^2+...$. 
In the DCDW state we then find $H_D=
=-i\balpha\cdot\nabla+\gamma^0[\frac{1+\gamma_5\tau_3}{2}M(\bx)+\frac{1-\gamma_5\tau_3}{2}M^*(\bx)]$ with $M(\bx)=m(\bx)\exp(i\theta(\bx))$. The phase degree of freedom $\theta(\bx)$ or the complex order parameter $M(\bx)$ then gives rise to important features. 
The 1+1 dimensional version of DCDW, 
called {\it chiral spiral}, has been studied in ref.\cite{dun}, where chiral anomaly and the nesting effect play important roles to establish chiral spiral: 
chiral anomaly gives baryon density as $\rho_B=\mu/\pi$ for chemical potential $\mu$, and the nesting effect $q=2\mu$ \cite{dun}. In particular it should be interesting 
to observe the latter relation is similar to the one in charge density wave or spin density wave in quasi-one dimensional system in condensed matter physics \cite{nes}. Consequently it has been shown that the chiral spiral is the most favorite configuration among various form of the condensate \cite{dun}. In 1+3 dimensions  anomalous relation $\rho_B=\mu/\pi$ becomes irrelevant and the nesting effect becomes incomplete. However, it has been shown that DCDW appears in the limited region of chemical 
potential \cite{nak}. 

Recently the chiral transition or deconfinement transition has attracted much attention in the presence of the magnetic field. The magnetic field is familiar in QCD through phenomena of compact stars \cite{comp} or high-energy heavy-ion collisions \cite{revmag}. 
Theoretically, SSB has been shown to be enhanced by the magnetic effect, sometimes called {\it magnetic catalysis}, and the chiral magnetic effect has been another interesting subject \cite{revmag}. Recently 
the lattice QCD simulations have 
started to explore the chiral phase digram on the temperature ($T$) -magnetic field ($B$) plane \cite{lat}. One of the great advantages may be then that it is free from the sign problem on this plane. 

Here we discuss DCDW in the presence of the uniform magnetic field, and thereby the chiral phase transition in the $\mu-T-B$ space. 
The energy levels of quarks then are discretized in the plane perpendicular to the magnetic field to produce the Landau levels and some 1+1 dimensional feature appears by way of the lowest Landau level (LLL); chiral anomaly revives through the spectral asymmetry of the Dirac operator.   
Here we demonstrate it by using the NJL-like model in the mean-field approximation, which is one of the effective models of QCD at low energy scale.
Consider the Dirac operator, $H_D=\balpha\cdot\bP+\gamma^0 \left[\frac{1+\gamma_5\tau_3}{2}M(\bx)+\frac{1-\gamma_5\tau_3}{2}M^*(\bx)\right]$ 
with $\bP=-i\nabla+Q{\bf A}$, where $Q={\rm diag}(2/3e,-1/3e)$ is the charge matrix. 
We take the direction of the magnetic field $\bf B$ along $z$ axis.
Consider for a while a single flavor by putting $\tau_3=1$ and $Q=\tilde e>0$, and take a generic form of $\theta(\bx)$. Changing the basis by the Weinberg transformation (local chiral $U(1)$), $\psi\rightarrow \psi_W=\exp(i\gamma_5\tau_3\theta(\bx)/2)\psi$, the Dirac operator can be written as 
\beq
{\tilde H}_D=\balpha\cdot\bP+\gamma_0 m(\bx)-\gamma_0\gamma_5\bgamma\nabla\theta(\bx)/2.
\eeq
Considering the flavor symmetric quark matter, $\mu_u=\mu_d(\equiv \mu)$, the quark number then can be generally given as
\beqa
\langle {\hat N}\rangle=-\frac{1}{2}\eta_H&+&\sum_k {\rm sign}(\lambda_k)\left[\theta(\lambda_k)n_F(\lambda_k-\mu)\right.\nonumber\\
&+&\left.\theta(-\lambda_k)(1-n_F(\lambda_k-\mu))\right],
\eeqa
where $\lambda_k$ is the eigenvalue of $H_D$ and $n_F(\omega)=(1+e^{\omega/T})^{-1}$ \cite{nie}. The first term is a topological quantity called the Atiyah-Patodi-Singer $\eta$ invariant \cite{aps},
\beq
\eta_H=\lim_{s\rightarrow 0+}\eta_H(s),~~~\eta_H(s)=\sum_k{\rm sign}(\lambda_k)|\lambda_k|^{-s}
\eeq
 and may take a non-vanishing value if the spectrum of $H_D$ is asymmetry about zero. The second one is the usual expression given by the Fermi-Dirac distribution function. Using the Mellin transform, $\eta_H(s)$ can be written as
\beq
\eta_H(s)=\frac{1}{\pi}\cos\left(\frac{s\pi}{2}\right)\int_0^\infty d\omega\omega^{-s}\int d^3x {\rm tr}
\left[R_E(\bx,i\omega)+{\rm c.c.}\right],
\label{reseta}
\eeq
where $R_E$ is the Euclidean resolvent,
\beq
R_E(\bx,i\omega)\equiv \langle \bx\left|\frac{1}{{\tilde H}_D-i\omega}\right | \bx \rangle=\langle \bx \left|\gamma_0S(i\omega)\right|\bx\rangle,
\label{res}
\eeq
with the propagator, $S(i\omega)$, 
$
S^{-1}(i\omega)=S_A^{-1}(i\omega)+\delta S
$
with  $\langle\bx\left|\delta S\right|\by\rangle=\gamma_5\bgamma\cdot\nabla\theta(\bx)/2\delta(\bx-\by)$.  $S_A$ is the Green's function in the presence of the magnetic field without DCDW. For slowly varying $\theta(\bx)$,  we can apply the adiabatic method of Goldstone and Wilczek \cite{gw}. We can approximate 
$m(\bx)=m+...$ in the lowest order. Writing $S_A(x,y)={\rm exp}(i{\tilde e}\int_y^xd\bx\cdot{\bf A}){\tilde S}_A(x-y)$, the Fourier transform of ${\tilde S}_A(x-y)$   
can be decomposed over the Landau levels \cite{cho},
\beq
{\tilde S}_A(k)=ie^{-\bk_\perp^2/({|\tilde e}B|)}\sum_{n=0}^{\infty}\frac{(-1)^n D_n({\tilde e}B,k)}{(k^0)^2-(k^3)^2-m^2-2|{\tilde e}B|n},
\eeq
with the denominator,
\beqa
D_n({\tilde e}B,k)&=&(k_0\gamma^0-k^3\gamma^3+m)\left[{\cal P}_-L_n^0\left(u\right)
-{\cal P}_+L_{n-1}^0\left(u\right)\right]\nonumber\\
&&+4(k^1\gamma^1+k^2\gamma^2)L^1_{n-1}\left(u\right),
\eeqa
with $u=2\bk_\perp^2/|{\tilde e}B|$, where ${\cal P}_{\pm}=(1\pm i\gamma^1\gamma^2{\rm sign}({\tilde e}B))$ is the spin projection operator, and $L_n^\alpha(x)$ the generalized Laguerre polynomial. 
Expanding ${\tilde S}(i\omega)$ around ${\tilde S}_A$,  
$
{\tilde S}(i\omega)={\tilde S}_A(i\omega)-{\tilde S}_A(i\omega)\delta S{\tilde S}_A(i\omega)+...,
$
we have  
\beq
{\rm tr}R_E(\bx,i\omega)=-{\tilde e}/(4\pi)m^2/(m^2+\omega^2)^{3/2}{\bf B}\cdot\nabla\theta(\bx)+....
\eeq
There are two remarks in order: only LLL contributes and the result include only the inner product of $\bf B$ and $\nabla\theta$. Substituting it into Eq.~(\ref{reseta}) we find 
\beq
\eta_H=\lim_{s\rightarrow 0}\eta_H(s)=-\frac{{\tilde e}}{2\pi^2}\int d^3x{\bf B}\cdot\nabla\theta(\bx)+....
\label{etainv}
\eeq
Thus the quark-number density can be written as 
\beq
\rho_B^{\rm anom}=\frac{{\tilde e}}{4\pi^2}{\bf B}\cdot\nabla\theta(\bx)+....
\eeq
This formula is the same as the one given by Son and Stephanov by gauging the Wess-Zumino-Witten action \cite{son}. 
Thus we find that the leading term in $\eta_H$ originates from chiral anomaly and model independent, 
while other terms are model dependent. 
Here it is interesting to observe that $\eta_H$ is independent of  the dynamical mass  $m$, which is one of the remarkable features of chiral anomaly. 
It should be worth mentioning that the anomolous baryon number has been evaluated in the chiral bag model for nucleon \cite{gol}: quarks inside the bag exhibit the spectral asymmetry, and the baryon number is then given by the sum of the quarks, skyrmion and the anomalous baryon number to be one.  
Since $\lambda_k$ changes its sign under the  $CT$ transformation, $\psi\rightarrow i\gamma_0\gamma_5\psi$, $\lambda_k(M)\rightarrow -\lambda_k(M^*)$, we can see $\eta_H$ always vanishes for real order parameter: the spectrum of the Dirac operator is symmetric about the zero eigenvalue for $M\in {\bf R}$. The phase degree of freedom $\theta(\bx)$ is important in our case.

Accordingly, the thermodynamic potential should includes the anomalous term besides the usual piece $\Omega_s$, $\Omega=\Omega_s+\Omega_{\rm anom}$.
By way of the thermodynamic relation, $\rho_B^{\rm anom}=-\partial\Omega_{\rm anom}/\partial\mu$, we have 
\beq
\Omega_{\rm anom}=-\frac{{\tilde e}\mu}{4\pi^2}\int d^3x {\bf B}\cdot\nabla\theta(\bx)+....
\label{therm}
\eeq
Taking $\theta(\bx)=\bq\cdot\bx$ for DCDW, we immediately find from Eq.~(\ref{therm}) that 
the most favorite direction of the wave vector $\bq$ is parallel to $\bf B$ in the weak magnetic field. The authors in ref.\cite{fro} have also found
that the effective energy increases by a small deviation from the parallel configuration. 

It should be interesting to see that the $\eta$ invariant can be directly evaluated in the closed form without recourse to the derivative expansion for the case,  ${\bf B}//{\bf q}$. Using the Landau gauge, ${\bf A}=(0,Bx,0)$, the Dirac operator $H_D$ can be reduced to $4\times 4$ matrix on the basis of the plane wave $\exp(ik_3z+ik_2y)$ and the Hermite functions $u_n(x)$ \cite{fro}, where $n$ specifies the Landau levels. However, for the lowest Landau level (LLL), $n=0$, $H_D$ is reduced to $2\times 2$ 
matrix from the property of $u_n(x)$.
Thus the energy spectrum of the Dirac Hamiltonian then can be obtained, 
\beqa
\lambda_{n,p,\zeta,\epsilon}&=&\epsilon\sqrt{\left(\zeta\sqrt{m^2+k_3^2}+q/2\right)^2+2eBn}, n=1,2,..., \nonumber\\ 
\lambda_{n=0,p,\epsilon}&=&\epsilon\sqrt{m^2+k_3^2}+q/2,~~~({\rm LLL}),
\eeqa 
with $\zeta=\pm 1,\epsilon=\pm 1$.
We can immediately see the spectrum is symmetric about zero except LLL: LLL exhibits spectral asymmetry due to the reduction of the Dirac operator. 
The evaluation of $\eta_H$ is straightforward in this case and results 
in the same value as (\ref{etainv}) without any higher-order term \cite{tato}. 

After taking $\bq$ along $\bf B$, we can see another implication of chiral anomaly. Since $\Omega_s$  is the even function of $\bq$, the minimum point of $|\bf q|$ is always shifted from zero. Thus we find the DCDW phase is favorite for $\mu\neq 0$ in the presence of the magnetic field, irrespective of the dynamical mass.
In the following we shall reveal another interesting aspect of chiral anomaly around the transition point, invoking the generalized Ginzburg-Landau (gGL) theory.

Consider the general expansion of the thermodynamic potential near the transition point,
\beqa
\Omega(M)&=&\Omega(0)
+\frac{\alpha_2}{2}\left|M\right|^2+\alpha_3{\rm Im}\left(MM'^*\right)\nonumber\\
&&+\frac{\alpha_{4a}}{4}\left|M\right|^4+\frac{\alpha_{4b}}{4}\left|M'\right|^2+...
\eeqa
with a shorthand notation $M'\equiv dM/dz$, where we used the property that $\Omega(M)$ is invariant under the global chiral transformation, $M\rightarrow e^{i\phi}M$.
If the Dirac operator is symmetric by exchanging $M(z)$ and $M^*(z)$, the imaginary terms are absent. 
DCDW in the absence of the magnetic field satisfies this condition, while it breaks in the presence of the magnetic field.
The coefficients $\alpha_n$ are functions of thermodynamic variables, $\mu,T,B$ \cite{nic,dun}.

In the absence of the magnetic field, the spectrum becomes symmetric about zero and the coefficient $\alpha_3(\mu,T,0)=0$. Thus the Lifshitz point is given 
by looking at the leading-order contributions \cite{lif},
$
\alpha_2(\mu,T,0)=\alpha_{4b}(\mu,T,0).
$
Within the NJL model, $\alpha_{4a}(\mu,T,0)=\alpha_{4b}(\mu,T,0)$, so that the Lifshitz point coincides with the tricritical point for the chiral transition with the uniform condensate \cite{nic}. We can see that $\alpha_3(\mu,T,B)$ becomes non vanishing in the presence of the magnetic field.
Thus gGL theory should bring about qualitatively different consequences. Most important and interesting one may be the appearance of the novel Lifshitz point. 
This point is defined as the tricritical  where the two lowest nontrivial coefficients vanish:
\beq
\alpha_2(\mu,T,B)=\alpha_3(\mu,T,B)=0,
\eeq
for given $B$. First we evaluate $\alpha_2(\mu,T,B)$ in the presence of the magnetic field in 1+3 dimensions, by using the 2 flavor NJL model. 
Since it includes divergence, we need some regularization. 
Applying then the proper-time regularization with cutoff $\Lambda$, we have
\beqa
&&\alpha_2(\mu,T,B)=
-\sum_{f;m\ge 0,n}\frac{N_c|e_fB|}{\pi^2}T(2-\delta_{n,0})\times\nonumber\\
&&\times{\rm Im}\int_{\Lambda^{-2}}^\infty d\tau\sqrt{\frac{\pi}{i\tau}}e^{i\tau[(\omega_m+i\mu)^2+2|e_fB|n]}+\frac{1}{2G} \nonumber\\   
\label{alpha2}
\eeqa
with the Matsubara frequency, $\omega_m=(2m+1)\pi T$, where we revive the flavor dependence by using $e_{f=u,d}$ instead of $\tilde e$. . 
 In particular, for $\mu=0$, the first term reads 
$
-4N_c\sum_{f;m\ge 0,n}\frac{|e_fB|}{(2\pi)^2}T\sqrt{\pi}\lambda_{m,n}^{-1}\Gamma\left(\frac{1}{2}, \frac{\lambda_{m,n}^2}{\Lambda^2}\right)
$
with $\lambda_{m,n}^2=\omega_m^2+2|e_fB|n$,  where $\Gamma(a,x)$ is the incomplete Gamma function. For $x\rightarrow \infty, |{\rm arg} x|<3\pi/2$, 
it behaves $\Gamma(a,x)=e^{-x}x^{a-1}[\sum_{n=0}^{N-1}(1-a)_n(-x)^{-n}+O(|x|^{-N})]$ \cite{mag}, so that $\alpha_2$ becomes finite. 

\begin{figure}[h]
\includegraphics[width=15pc]{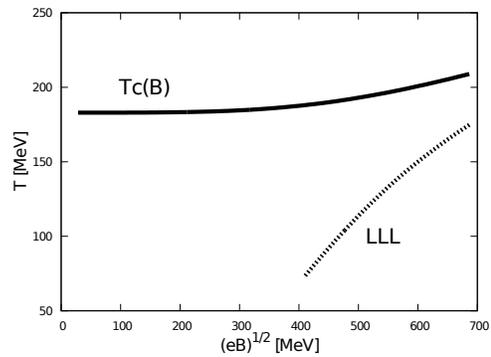}
\caption{\label{fig:03} Critical temperature (Lifshitz point) on the $\mu=0$ plane as a function of $B$. The same values are 
used for the parameters as in ref. \cite{nak}: $G\Lambda^2=6.35$ The dotted curve is given by using only LLL contribution, which 
indicates the dimensional reduction in large $B$.
}
\end{figure}

The coefficient $\alpha_3(\mu,T,B)$ includes no divergence. To evaluate $\alpha_3(\mu,T,B)$ 
it should be sufficient to consider the LLL contribution,
\beqa
&&\alpha_3(\mu,T,B)=-\sum_f\frac{N_c|e_fB|}{16\pi^3T}{\rm Im}\psi^{(1)}\left(\frac{1}{2}+i\frac{\mu}{2\pi T}\right).~~~~~~~~
\eeqa
since other contributions vanish, where $\psi^{(1)}$ is the trigamma function. Note that $\alpha_3(\mu,T,B)\ge 0$.
Then $\alpha_3(\mu,T,B)=0$ implies $\mu=0$: the Lifshitz point resides on this plane. 
Note that this result does not depend on the details of the model, but comes from chiral anomaly: 
vanishing of chiral anomaly simply means $\mu=0$. In Fig.~1 we show the Lifshitz line on the $B-T$ plane, determined by 
the equation, $\alpha_2(0,T,B)=0$. Note that the critical temperature increases as $B$ does in our calculation, 
while the recent lattice QCD simulation has suggested its decrease \cite{lat}. However, our conclusion of 
the Lifshitz point on the $B-T$ plane should hold irrespective of its tendency on $B$.

We have shown that the Lifshitz point for the inhomogeneous chiral phase should reside on $B-T$ plane given by $\mu=0$. 
This conclusion may be model-independent and lead by chiral anomaly.
A clear evidence may be obtained for small $\mu$, where the wave vector is proportional to the strength of the 
magnetic field $B$ and chemical potential $\mu$. 
For $\alpha_{4a,b}(\mu,T,B)>0$, the optimum values of the amplitude $m$ and wave vector $q$ are determined by the conditions 
$\partial\Omega/\partial m=\partial\Omega/\partial q=0$, and we find $q=-2\alpha_3(\mu,T,B)/\alpha_{4b}(\mu,T,B)$. 
Since $\alpha_3(\mu,T,B)$ should be proportional to $\mu B$, $q$ is as well.
The critical line on the $\mu-T$ plane, where the amplitude vanishes but wave vector necessarily does not, is given by the equation, 
$\alpha_2(\mu,T,B)\alpha_{4b}(\mu,T,B)=4\alpha_3^2(\mu,T,B)$ for given $B$. The critical line is then shifted upward 
from the usual chiral transition given by the uniform condensate, $\alpha_2(\mu,T,B)=0$, assuming SSB at low $T$ and small $\mu$ 
(see Fig.~2 for example). 
Since $\mu\simeq 0$ region is free from the sign problem, this critical line can be examined by 
the lattice QCD simulation. 

\begin{figure}[h]
\includegraphics[width=11pc]{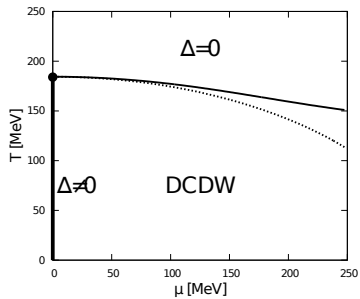}
\caption{\label{fig:03} Phase diagram in the $\mu-T$ plane near the Lifshitz point for $(eB)^{1/2}\simeq 300{\rm MeV}$, 
where we approximate $\alpha_{4b}(\mu,T,B)$ and $\alpha_2(\mu,T,B)$ by their values at $B=0$. 
The uniform phase with $\Delta\neq 0$ is confined on the $\mu=0$ axis (bold line). 
Solid line shows the boundary between the DCDW phase and the chiral restored phase ($\Delta=0$), while dotted line corresponds to 
the usual chiral transition. 
}
\end{figure}

In the presence of the small current mass for quarks the chiral transition becomes cross-over for the usual chiral transition, 
but the Lifshitz point should survive and we have a clear second-order phase transition from the inhomogeneous chiral phase to the uniform 
phase even in this case \cite{kar}.

Finally we briefly discuss the relation of DCDW with RKC in the presence of the magnetic field, leaving full discussion in another paper \cite{nis}. 
Considering the hybrid condensate, 
$
M(z)=m\left(\frac{2\sqrt{\nu}}{1+\sqrt{\nu}}\right){\rm sn}\left(\frac{2m z}{1+\sqrt{\nu}}; \nu\right)\exp(iqz),
$ 
we can discuss two phases simultaneously, where ${\rm sn}(x;\nu)$ is the Jacobian elliptic function with modulus $\nu$.
One can easily check this is one of the Hartree-Fock solutions in the 1+1 dimensional NJL$_2$ model.
We can immediately see that the anomalous term arises in the thermodynamic potential from the wave vector $q$ even in this case. 
Hence the non-vanishing $q$ is always favorite and pure RKC phase never appears in the presence of the magnetic field. 

 \begin{acknowledgments}
We thank H. Abuki and R. Yoshiike for useful discussions. This work is partially supported by Grants-in-Aid for Scientific Research on Innovative Areas 
through No. 24105008 provided by MEXT.
\end{acknowledgments}

\end{document}